\documentclass[sigconf]{acmart}
\usepackage{multirow}    

\AtBeginDocument{%
  \providecommand\BibTeX{{%
    \normalfont B\kern-0.5em{\scshape i\kern-0.25em b}\kern-0.8em\TeX}}}

\setcopyright{acmcopyright}
\copyrightyear{2022}
\acmYear{2022}
\acmDOI{10.1145/1122445.1122456}

\copyrightyear{2022}
\acmYear{2022}
\setcopyright{acmcopyright}\acmConference[LAK22]{LAK22: 12th International Learning Analytics and Knowledge Conference}{March 21--25, 2022}{Online, USA} \acmBooktitle{LAK22: 12th International Learning Analytics and Knowledge Conference (LAK22), March 21--25, 2022, Online, USA}
\acmPrice{15.00}
\acmDOI{10.1145/3506860.3506874}
\acmISBN{978-1-4503-9573-1/22/03}

\begin{document}


\title[How can Email Interventions Increase Students' Completion of Online Homework?]{How can Email Interventions Increase Students' Completion of Online Homework? A Case Study Using A/B Comparisons}


\author{Angela Zavaleta-Bernuy}
\affiliation{\institution{University of Toronto}
\country{Canada}}
\email{angelazb@cs.toronto.edu}

\author{Ziwen Han}
\affiliation{\institution{University of Toronto}
\country{Canada}}
\email{ziwen.han@mail.utoronto.ca}

\author{Hammad Shaikh}
\affiliation{\institution{University of Toronto}
\country{Canada}}
\email{hammy.shaikh@mail.utoronto.ca}

\author{Qi Yin Zheng}
\affiliation{\institution{University of Toronto}
\country{Canada}}
\email{qiyin.zheng@mail.utoronto.ca}

\author{Lisa-Angelique Lim}
\affiliation{\institution{University of Technology Sydney}
    \country{Australia}
}
\email{lisa-angelique.lim@uts.edu.au}

\author{Anna Rafferty}
\affiliation{\institution{Carleton College}
    \country{United States}
}
\email{arafferty@carleton.edu}

\author{Andrew Petersen}
\affiliation{
    \institution{University of Toronto Mississauga}
    \country{Canada}
}
\email{andrew.petersen@utoronto.ca}

\author{Joseph Jay Williams}
\affiliation{\institution{University of Toronto}
    \country{Canada}}
\email{williams@cs.toronto.edu}

\renewcommand{\shortauthors}{Zavaleta-Bernuy, et al.}
\newcommand{\ignore}[1]{}


\begin{abstract}


\noindent Email communication between instructors and students is ubiquitous, and it could be valuable to explore ways of testing out how to make email messages more impactful. This paper explores the design space of using emails to get students to plan and reflect on starting weekly homework earlier. We deployed a series of email reminders using randomized A/B comparisons to test alternative factors in the design of these emails, providing examples of an experimental paradigm and metrics for a broader range of interventions. We also surveyed and interviewed instructors and students to compare their predictions about the effectiveness of the reminders with their actual impact. We present our results on which seemingly obvious predictions about effective emails are not borne out, despite there being evidence for further exploring these interventions, as they can sometimes motivate students to attempt their homework more often.  We also present qualitative evidence about student opinions and behaviours after receiving the emails, to guide further interventions. These findings provide insight into how to use randomized A/B comparisons in everyday channels such as emails, to provide empirical evidence to test our beliefs about the effectiveness of alternative design choices. 


\ignore{
\textbf{Motivation} Instructors do not always know the best ways to communicate with students through email, but they might have an intuition of the most effective option. How can instructors test their intuitions and collect feedback on them?\\
\textbf{Objectives} This study provides a guide for instructors wishing to test their design choices and to demonstrate how their predictions might or might not work in particular contexts.\\
\textbf{Methods} We deployed a series of email reminders using A/B experiments. We also surveyed and interviewed instructors and students to compare their predictions about the effectiveness of the reminders with their actual impact.\\
\textbf{Results} We find that email reminders motivate students to start the homework earlier regardless of the version of the email tested. We also present qualitative evidence  that instructor intuition might not always predict the most optimal condition for students.\\
\textbf{Discussion} These findings highlight the importance of using A/B comparisons as everyday tools to provide empirical evidence to improve design choices.
}
\end{abstract}


\maketitle

\section{Introduction}

The ability to hold and sustain student interest and engagement in course learning activities is a persistent challenge for educators. In particular, studies of online courses have evidence of issues related to high withdrawal rates and low levels of course evaluation and satisfaction when compared to more traditional on-campus-based offerings \cite{bawa2016retention}. As well, student participation in learning activities has been shown to be a significant factor influencing academic performance in online courses \cite{finn2012student, you2016identifying}. Thus, educators are faced with the problem of developing economical yet scalable practices that can assist students in sustaining their engagement in learning activities, especially in online learning modes.  Many instructors send emails and announcements to contact students. How can we use emails as a channel for helping students stay on track with their homework, either online or in-person? In other contexts, reflecting and making plans about the next steps improves the chances that people will follow through with their goals \cite{rodriguez2018students, VANDENBOOM2004551}. In this paper, we explore different ways of using emailed homework reminders to prompt students to take steps that might increase students' engagement with the homework. 

Past work has identified situations in which particular interventions can be promising-- like particular implementations of reflection prompts \cite{VANDENBOOM2004551} and reminders \cite{edwards2015examining} -- although in other contexts similar interventions do not have an effect \cite{baker2016}, underscoring the importance of empirically testing potential alternative implementations in a context. We explored building alternative versions of emails to students by combining various components: The subject line prompts to spend a few minutes planning, reading homework, or thinking of advice to others to start early, links to external sources or learning resources. Given the broad and complex space of emails that instructors and researchers can write, we design and deploy randomized A/B comparisons (also called randomized controlled trials/experiments) that compare alternative emails, providing examples of how to test the impact and perception of alternative ideas for how to help students keep engaged. 

We quantified the potential impact on students by looking at students’ academic performance such as homework grades, attempt rate, and start time. We collected qualitative data from students about the advantages and disadvantages of various design choices, how these reminders fit into their lives and how these reminders are received \cite{kizilcec2017diverse, lim2021impact}. We also collected qualitative data from instructors to contrast their predictions of the impact of various design choices with the actual impact observed. We found that many intuitive interventions predicted to have an impact (e.g. emails from instructors vs. teaching assistants) did not change student behaviour. However, we did find that on average, repeated emails did have an impact over an extended period of time in getting students to attempt homework more often (generalizing past work where we saw only an immediate benefit \cite{zavaleta2021investigating}). Our broader goal is for the details of the experimental design process in these case studies to provide insight into how instructors and researchers can design, deploy, and quantitatively and qualitatively evaluate a wide range of alternative approaches to messaging students. 





\section{Related Work}

\textbf{Embedded experimentation:} Motz et al. highlighted the importance of scalable embedded experimentation in providing an accurate characterization of education in real-world scenarios, while also being able to make causal inferences. They also highlighted the necessity of working with instructors in the process of designing experimentation to further the applicability of results \cite{motz2018causal}. Our study addresses this framework through the deployment of A/B comparisons to isolate the effect of deliberate design factors. Furthermore, our case study demonstrates how to instantiate this form of embedded experimentation into practice. Though the involvement of instructors is important, we also seek to address a quantitative approach to evaluating instructor intuitions by comparing predictions to data collected.

\textbf{Student engagement and study habits:} Procrastination is one of the biggest barriers to academic performance \cite{yilmaz2017relation}. Procrastination can be reduced by student engagement in class and a positive learning environment \cite{katz2014ll}. Through student and instructor surveys, Martin and Bollieger found that instructors want to increase students' engagement, inside and outside class, because it motivates them to learn and improves their performance \cite{martin2018engagement}. More recently, Goodsett conducted a literature review about the best teaching practices in online learning and found that as online learning can be more self-paced compared to in-class delivery, instructors should consistently motivate students and engage them in active learning \cite{goodsett2020best}. Previous studies also show the importance of improving self-regulated learning strategies in flipped classrooms, and how it can be achieved with goal-setting skills and reflection prompts \cite{falkner2014identifying, schwarzenberg2020supporting, VANDENBOOM2004551}. Van den Boom et al. investigated the effects of reflection prompts in distance education and web-based learning environments and found that they encouraged students to take action on a particular task
\cite{VANDENBOOM2004551}. Similarly, Rodriguez et al. used self-reporting and self-testing practices (i.e. self-evaluation through practice problems) to improve study habits, which they found to be correlated with course grades \cite{rodriguez2018students}.

\textbf{Mitigating procrastination:} In online courses, reflection prompts that provide self-regulated learning scaffolds, like motivation and time management, have a human factor that encourages closer student-instructor interaction \cite{effects2020}. Song and Kim concluded that the human factor is helpful in self-regulated learning and course performance, encouraging students to participate in the course \cite{effects2020}. Azfal and Jami recommended that university instructors explore different possibilities to promote time management and self-regulation skills to increase student motivation for homework completion \cite{afzal2018prevalence}. Klassen et al. recommended that building students’ confidence is also an efficient way to address procrastination after assessing students' responses from the Motivated Strategies for Learning Questionnaire \cite{klassen2008academic}.

The benefits of addressing procrastination and suggesting different study strategies in online courses are also reflected in students' performance \cite{michinov2011procrastination, hartwig2012study}. Michinov et al. evaluated student time management and performance in online learning and provided empirical evidence of the negative relationship between procrastination and performance as students who are high procrastinators tend to engage less with the course material in general \cite{michinov2011procrastination}. Moreover, in a study conducted by Hartwig and Dunlosky in a large undergraduate course that evaluated students' study strategies, those who reported scheduling their work in advance had higher grade point average (GPA) in their overall university academic performance than students who had trouble managing deadlines \cite{hartwig2012study}.

\textbf{Reminder messages:} Researchers have deployed similar interventions to the one described in this paper, with the goal of reducing procrastination by sending email and text messages reminders. Edwards et al. sent students an email with their previous homework performance, which caused significantly reduced rates of late submissions and increased rates of early submissions \cite{edwards2015examining}. Humphry et al. focused on the format of the reminders that students received and used randomized A/B comparisons to analyze the effect of receiving an email reminder compared to a text message \cite{humphrey2019exploring}. The results of this study showed that text reminders had a positive impact on on-time assignment submissions and overall performance in the class \cite{humphrey2019exploring}. In a Massive Open Online Course (MOOC), Baker et al. prompted students to schedule a reminder nudge about their course videos to measure their engagement with the course material \cite{baker2016}. Although they found no impact on student attitudes in their settings, the authors suggested that these type of interventions would have success in a context where students are highly motivated by course credits \cite{baker2016}. 

The research by Nikolayeva et al. is particularly relevant for the current study \cite{nikolayeva2020does}. They also conducted randomized control trials to test out four different versions of email reminders. Their study found that simple, non-personalized email reminders to complete optional online quiz exercises in a blended Physics course had a greater impact than longer, more personalized emails that specifically prompted students’ time management goals. While their study contributed to research on effective reminder emails for self-regulation with respect to time management, it stopped short of providing important information about the reasons for this effect. As noted by the researchers, they did not solicit students’ reactions to the email content - this would have provided greater insights into the quantitative results, especially, how the more sophisticated versions of reminders were perceived by students. Moreover, the email reminders targeted optional quizzes rather than required course tasks; as noted by the authors, it would be interesting to investigate how different email reminders would affect students’ completion of course tasks that were assessed, which is the goal of our current study. Our study also tests a broader set of A/B comparisons over multiple weeks to maximize the number of comparisons we make and to evaluate effectiveness at specific time points. This helps narrow down conditions which have the potential for further exploration. In addition, we make use of homework start time as a more direct measurement of the effect conditions have in reducing student procrastination.

Even though work has been done to investigate students’ opinions about their behaviour towards online courses including homework performance and procrastination \cite{zheng2016ask}, and evidence about their course performance after receiving reminders, to our knowledge, there is not much work that focuses on students' perceptions of those reminders, or aims evaluate the difference between instructor intuition and quantitative results. 

Our past work \cite{zavaleta2021investigating}, which analyzed students' attempt and completion rate for homework during a single week of the four weeks of intervention described in this paper, provided initial evidence that reminder messages may have motivated some students who would otherwise not have attempted the homework to at least start. The current study aims to extend these findings over a longer time frame. Our study was guided by the following research questions: 1) What is the overall impact of reminder messages on homework performance? 2) What is the impact of reminder design on homework performance? 3) What are students’ perceptions of reminder messages? 4) How does instructor intuition of the reminder messages differ from observed results?

\section{Methods}

To test the reminder messages, we use A/B comparisons which have been used in similar experiments \cite{humphrey2019exploring, necamp2019beyond, yeomans2017planning, o2018can}, to find the most effective option from different factors \cite{williams2018enhancing}. A/B comparisons, also known as randomized control trials (RCT), are a common practice used in different areas to find the most effective option from different variations \cite{williams2018enhancing}. In this paper, we show how A/B comparisons for emails reminders are a low-cost and effective strategy to test multiple messages alternatives and find the one that works better for students, without the use of external tools or resources.

\subsection{Course Setting}

We deployed the reminder messages interventions during the second semester of the academic year in an introduction to Computer Science course (CS1) in a research-focused North American university. Our goal was to help students start earlier on the weekly graded homework. These multiple-choice and open-ended programming questions about Python were completed before the lecture (worth 0.5\% of their total marks and due on Mondays at 10 am) and after the lecture (worth 2\% of their total marks and due on Fridays at 5 pm), as part of a flipped/inverted classroom course. Each set of questions was released three to four days before the deadline and was due on the same day and time every week. The homework problems were submitted on the course's online homework platform, which provides immediate feedback to students. Students are allowed unlimited attempts for each question, with the best attempt being scored.
The reminder messages were deployed using the university’s Learning Management System, Canvas, which allows students to receive these reminders through its inbox and, by default, to receive an email notification. There were 1018 students enrolled in the course at the time of the interventions, 946 of whom provided explicit consent via a Qualtrics form and were included in the analysis. 

\begin{figure*}[htb]
    \centering
    \includegraphics[scale=0.45]{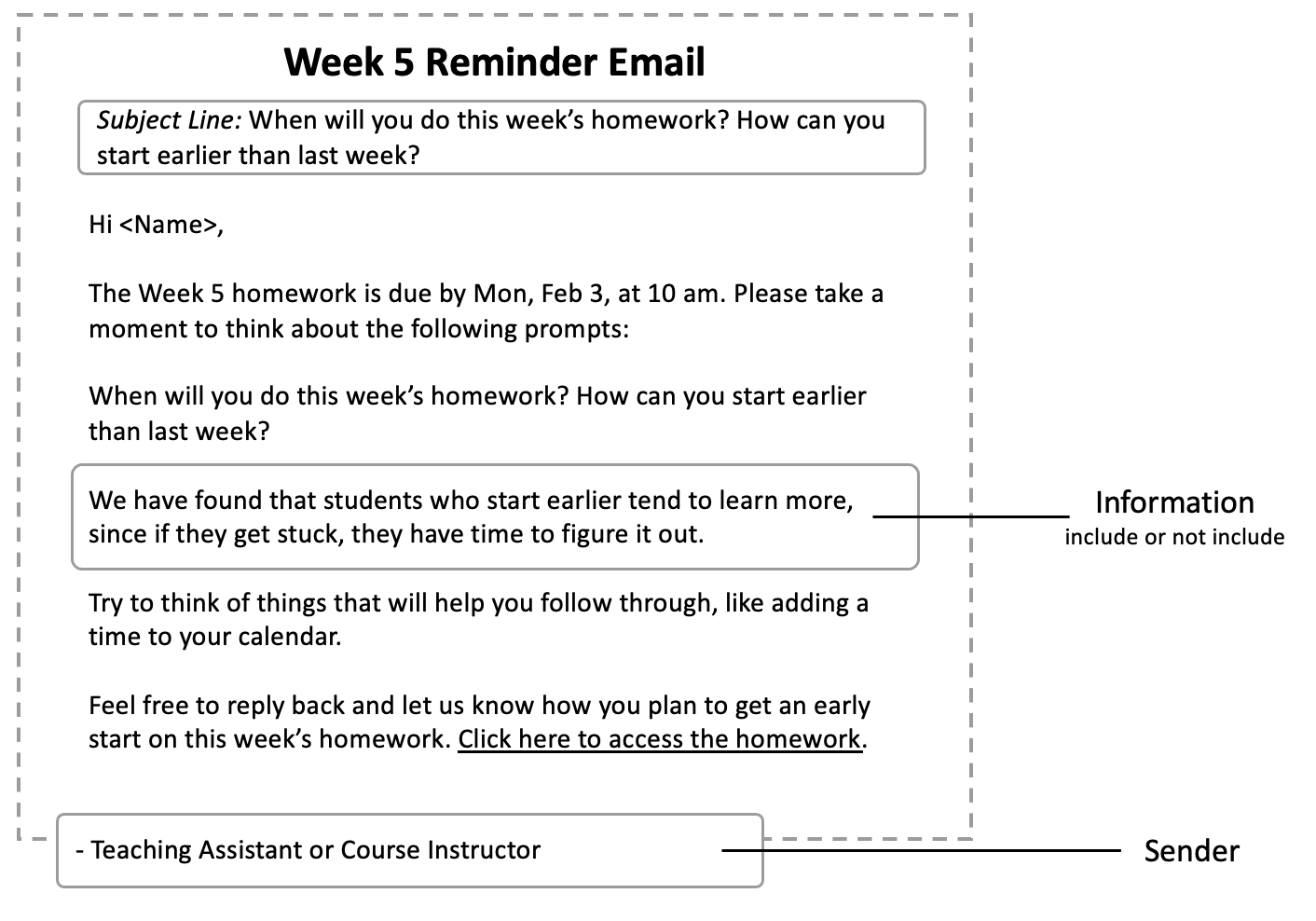}
    \caption{Example of reminder email sent during Week 5. Two variations were sent that week: sender and information content.}
    \vspace*{-1.5em}
    \label{fig:email-template}
\end{figure*}

\subsection{Data Collection}

Every week we randomized participants to the treatment group (receiving a reminder message through the Learning Management System with an email notification) or to the control group (a no-reminder message condition). This was double-blind in that neither instructors nor students knew which condition they were assigned to. We report on a sequence of interventions from weeks 5, 6, 7, and 8 in a 12-week course. The interventions started in week 5 as it was a point in the middle of the semester when students begin to fall behind as the course content becomes more challenging. 

We collected time-stamped homework submissions data from the online homework platform. We then compared metrics constructed from this data across the control and treatment groups. To illustrate our findings in students’ homework participation, we discuss the four weeks of interventions to show students’ responses to the emails throughout the term. The weeks 5 and 8 deployments targeted the pre-lecture homework assignment, while the weeks 6 and 7 deployments were reminders for the post-lecture homework assignment. The average send time of the reminder messages was 84, 51, 40, and 47 hours before the deadline respectively for weeks 5, 6, 7, and 8. The week 5 and week 8 homework was posted 96 hours before the deadline,  week 6 and 7 homework was 149 hours before the deadline. 

Our goal was to measure the impact of reminder messages on student behaviour. Accordingly, we measured behavior changes across the following outcomes: 1) the number of hours before a deadline that students started homework (start time), and 2) homework attempt rate. The reminder messages intended to encourage students to start homework earlier by getting them to set a plan or encouraging them to begin. However, one could also imagine these reminder messages would not change students' behaviors as students would not find them useful or would simply ignore the messages. 

We investigate this question by looking at the impact on students' starting time relative to the deadline, as well as exploring whether the emails might increase the tendency of students to even start the homework. The data for these analyses were collected directly from the online homework system and included the time at which the first problem was attempted (start time), the proportion who attempted at least one problem (attempt rate), and homework scores.

To understand students' perceptions and experience of receiving these messages, and what students found helpful or not about the reminders, we also conducted a post-intervention survey where we asked students four 7-point Likert scale questions to rate the helpfulness of the reminder messages. We also asked students to answer three open-ended questions about their feelings and the actions they took when receiving messages such as what they liked or did not liked about the reminders. The open-ended responses were manually coded by three researchers using thematic analysis to capture the insights about the intervention. While 935 students provided consent and responded to the Likert-scale questions, only 474 students answered the open-ended questions. We complemented this large collection of written responses with semi-structured interviews of a smaller sample of students (N=9) that probed similar questions in more depth. This sample of students was recruited by announcing the call for interviews after the course was completed, to only students who expressed interest in being contacted for a follow-up during the survey. 

To evaluate educator perspectives on the A/B designs, we surveyed two computer science instructors who have been regularly teaching and coordinating introductory courses for the past 10 years to predict the effectiveness of the reminder emails. Instructors were not shown any data or analysis. We separately presented them with the two conditions for each variation as described in Section 3.3 below and asked which would be better to send in terms of changing student behaviour in terms of (a) getting students to start earlier, and (b) increasing the rate of students doing the homework after receiving the email. We also tasked instructors with giving a quantitative prediction of attention rate (percentage of students paying attention to the emails) and action rate (percentage of students who as a result of the message start their homework early) changes between conditions after students receive the email. Instructors were finally asked to rate their confidence in their predictions. 

\begin{table*}
\begin{tabular}{l|l|cccc}
\hline
\multicolumn{2}{c}{Condition}                                                                                                                                                                                                                                                                     & \multicolumn{1}{l}{W5} & \multicolumn{1}{l}{W6} & \multicolumn{1}{l}{W7} & \multicolumn{1}{l}{W8} \\ \hline
\multirow{2}{*}{Sender}       & Instructor                                                                                                                                                                                                                                                        & \multirow{2}{*}{X}     & \multirow{2}{*}{}      & \multirow{2}{*}{}      & \multirow{2}{*}{}      \\ \cline{2-2}
                              & Teaching Assistant                                                                                                                                                                                                                                                &                        &                        &                        &                        \\ \hline
\multirow{3}{*}{Information}  & No prompt                                                                                                                                                                                                                                                         & \multirow{3}{*}{X}     & \multirow{3}{*}{}      & \multirow{3}{*}{}      & \multirow{3}{*}{}      \\ \cline{2-2}
                              & \multirow{2}{*}{\begin{tabular}[c]{@{}l@{}}"We have found that students who start earlier tend to learn more,\\ since if they get stuck, they have time to figure it out."\end{tabular}}                                                                          &                        &                        &                        &                        \\
                              &                                                                                                                                                                                                                                                                   &                        &                        &                        &                        \\ \hline
\multirow{2}{*}{Subject Line} & "When did you plan to do PCRS Perform? Can you start earlier?"                                                                                                                                                                                           & \multirow{2}{*}{}      & \multirow{2}{*}{X}     & \multirow{2}{*}{}      & \multirow{2}{*}{}      \\ \cline{2-2}
                              & "Remember to start early on PCRS Perform and finish before Fri at 5 pm"                                                                                                                                                                                           &                        &                        &                        &                        \\ \hline
\multirow{2}{*}{Recommendation}          & No prompt                                                                                                                                                                                                                                                         & \multirow{2}{*}{}      & \multirow{2}{*}{}      & \multirow{2}{*}{X}     & \multirow{2}{*}{}      \\ \cline{2-2}
                              & \begin{tabular}[c]{@{}l@{}}"One tip for starting is to just spend 5 minutes on it. This can help you\\ overcome the procrastination barrier, and make it less likely you forget. "\end{tabular}                                                                   &                        &                        &                        &                        \\ \hline
\multirow{4}{*}{Advice}       & No prompt                                                                                                                                                                                                                                                         & \multirow{4}{*}{}      & \multirow{4}{*}{}      & \multirow{4}{*}{}      & \multirow{4}{*}{X}     \\ \cline{2-2}
                              & \multirow{3}{*}{\begin{tabular}[c]{@{}l@{}}"Take a moment to think about the following prompt: What advice would\\ you give to other students to try to start earlier? Take 30 seconds to just\\ say out loud to yourself what you would tell them"\end{tabular}} &                        &                        &                        &                        \\
                              &                                                                                                                                                                                                                                                                   &                        &                        &                        &                        \\ 
                              \\
\end{tabular}%
\caption{Details of reminder email conditions for each variation. X indicates the week we tested the variation specified in each row.}
\label{tab:design_factors}
\end{table*}

\subsection{Reminder Message Design}

We designed the reminder messages using a broad set of models from work on the implementation intentions theory \cite{gollwitzer1999implementation} and with an emphasis on self-efficacy \cite{klassen2008academic}. Implementation intentions involve explicitly stating a goal that supports the transformation of good intentions into concrete behaviour change \cite{gollwitzer1999implementation}. Implementation intentions have been effective in self-regulatory strategies to bridge the gap between intentions and actions \cite{gollwitzer2006implementation, falkner2014identifying}, in improving homework performance by predicting grades \cite{sommer2012influences}, and habitual behaviour \cite{arbuthnott2009education}. The reminder messages used for these interventions encourage students to set a goal for when they will work on their homework and including students' past experiences on starting early and what they found useful. The reminders also contained a link to the homework problems. We created different versions of reminder messages which are similar in design with variations such as sender, subject line, and content (information, recommendation, and advice) presented to the students [Table \ref{tab:design_factors}]. We performed two manipulations in the first round of deployments, as we originally intended to explore the design space and potential impact of the emails. We tested one variation within the email content (information) and one external variation that involved the delivery method (sender). For the remaining deployments, we analyzed one new variation per email to focus on the prompts and nudges. An example of a full reminder email can be found in Figure \ref{fig:email-template}.

\subsubsection{Sender}

For Week 5, we randomized the sender of the reminder message. Students either received the reminder message from one of the course instructors or one of the teaching assistants [Table \ref{tab:design_factors}]. Moreover, the course instructor's learning management system's account had a profile picture associated with it for easy recognition, whereas the teaching assistant did not have a profile picture. We decided to randomize the sender to better understand the impact of authority on student behaviour and motivation. Receiving a message directly from an instructor can make students feel that they care about their learning and feel encouraged to engage with the course material \cite{goetz2021getting}. On the other hand, receiving a message from a teaching assistant to who students might be able to relate more, as they are students as well and could take them as peer role models, might make students feel the advice is more genuine \cite{clarke2018near}. 

\subsubsection{Information}

For week 5, we also randomized if students would receive information about the impact of starting homework early [Table \ref{tab:design_factors}]. The rush and urgency of doing homework at the last minute can cause poor performance and harm long-term retention \cite{you2015examining}. Presenting students a piece of information stating why starting homework early can help them may encourage them to do so. However, it is also possible that the precise reason given may not resonate with the student, leading the information to reduce the attention that they give to the reminder.

\subsubsection{Subject Line}

For week 6, we randomized the type of subject line of the reminder message [Table \ref{tab:design_factors}]. We are aware that not all students will open and read the reminder message, and therefore, we want to get students to take action even if they only pay attention to the subject line. We decided to randomize between two types of subject lines: a reminder deadline summary and a prompt to plan their homework. Showing students the deadline their homework and reminding them to start early can give them a sense of urgency that the due date is coming up and they should get to do the work. However, giving students a prompt to plan when they will do their homework and asking them if they can start earlier may allow students to think about the actions they need to take to achieve this goal \cite{rogers2015beyond}. Prompting people to make concrete plans increases the chances they will act on them and complete their tasks \cite{rogers2015beyond}.

\subsubsection{Recommendation}

For week 7, we randomized if students would receive a recommendation on how to start the homework early [Table \ref{tab:design_factors}]. Even though it might sound trivial how to start the homework, one of the hardest steps to complete a difficult task is starting, or even finding the motivation to start. With this tip, we recommend students to spend a few minutes looking at the homework to help them visualize what they need to complete. Moreover, following the implementation intentions theory, we present to students a concrete action that they can take in order to achieve the goal of starting early \cite{gollwitzer1999implementation}.

\subsubsection{Advice}

For week 8, we randomized whether students would receive a prompt asking them to reflect on the advice they would give to another student if they are trying to start their homework early [Table \ref{tab:design_factors}]. It may be easier for students to give recommendations to peers, and it can be difficult to recognize who is in the same spot as you \cite{kross2012boosting}. With this prompt, we want students to think about their classmates and what can help them, and also to realize that other students might be facing similar struggles. This prompt encourages collaboration and the social aspect of learning \cite{laal2012benefits}.





\subsection{Student Assignment}

We randomized whether or not students received reminder messages, to investigate the impact on their homework start times, as well as whether or not they started the homework. We also randomized which version of reminder email students in the treatment group received, as we tested the different variations of our factorial design. 

To evaluate the impact of the reminder messages, we conducted randomized A/B comparisons where only a portion of the class received the messages in any given week. For the first week, 2/3 of the students were assigned to the treatment group and received a reminder message; for the following three weeks the treatment group was re-randomized and extended to 3/4 of students so that more students could benefit from the reminder message after analyzing the first week's results. However, due to technical issues, 1/4 of the treatment group in week 7 did not receive a reminder message. The randomization of the control and treatment groups was done via Python's random shuffle function.





\begin{table*}[htbp]\centering
\def\sym#1{\ifmmode^{#1}\else\(^{#1}\)\fi}

\begin{tabular}{l|c c c c c c}
            &\multicolumn{1}{c}{(1)}&\multicolumn{1}{c}{(2)}&\multicolumn{1}{c}{(3)}&\multicolumn{1}{c}{(4)}&\multicolumn{1}{c}{(5)}&\multicolumn{1}{c}{(6)}\\
            &Started&Started&Homework&Homework&Start&Start\\
            & Homework & Homework &Performance &Performance & Time & Time\\
\hline
\multirow{2}{*}{Student Received Message}&      0.0465\sym{**}&      0.0371\sym{**}& 0.0508\sym{**} & 0.0388\sym{**}&      1.623  &     1.105         \\
            &    (0.0124)         &    (0.0130)    & (0.0135)&  (0.0138)   &    (1.046)         &    (1.062)         \\
\hline
Weekly Effects&         Yes         &         Yes  & Yes         &         Yes        &         Yes         &         Yes         \\
Participant Effects&          No         &         Yes         &          No         &         Yes &          No         &         Yes         \\
No. of students $\times$ week obs&        3784         &        3784         &        3784         &        3784  &        3234         &        3234        \\
\end{tabular}
\caption{Table entries representing the overall effects of receiving an email reminder on the propensity to start and complete homework. Results were obtained using a panel data regression on data from weeks 5 - 8. The outcome variable in Columns 1 and 2 is an indicator variable representing whether the student started the homework. For Columns 3 and 4, the outcome is an indicator variable representing the homework performance average. For Columns 5 and 6, the outcome is the hours before the deadline student started homework. The regressor in each scenario is an indicator of whether a student received an email. Rows 2 and 3 present whether week fixed effects and participant randomized effects were included in the regression model. The standard error of the treatment effect estimate is presented in parenthesis, and stars denote statistical significance: * \textit{p} < .05, ** \textit{p} < .01, \textit{z}-test.}
\label{tab:impact_rate}
\end{table*}

\section{Results}

\subsection{Quantitative Results}
\subsubsection{Overall Impact of Reminder Messages on Homework Performance}

\begin{figure}[htb]
    \centering
    \includegraphics[scale=0.45]{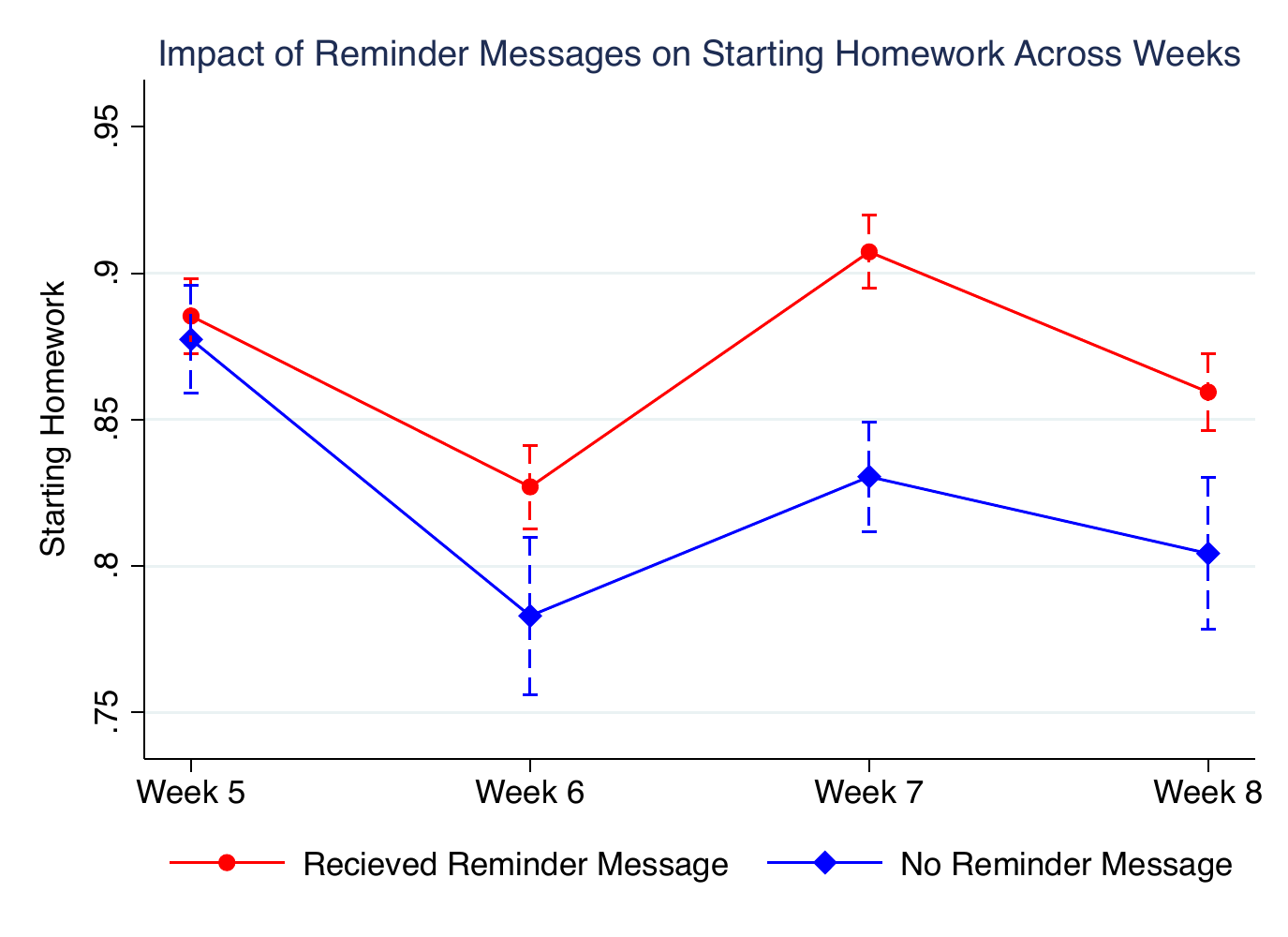}
    \includegraphics[scale=0.45]{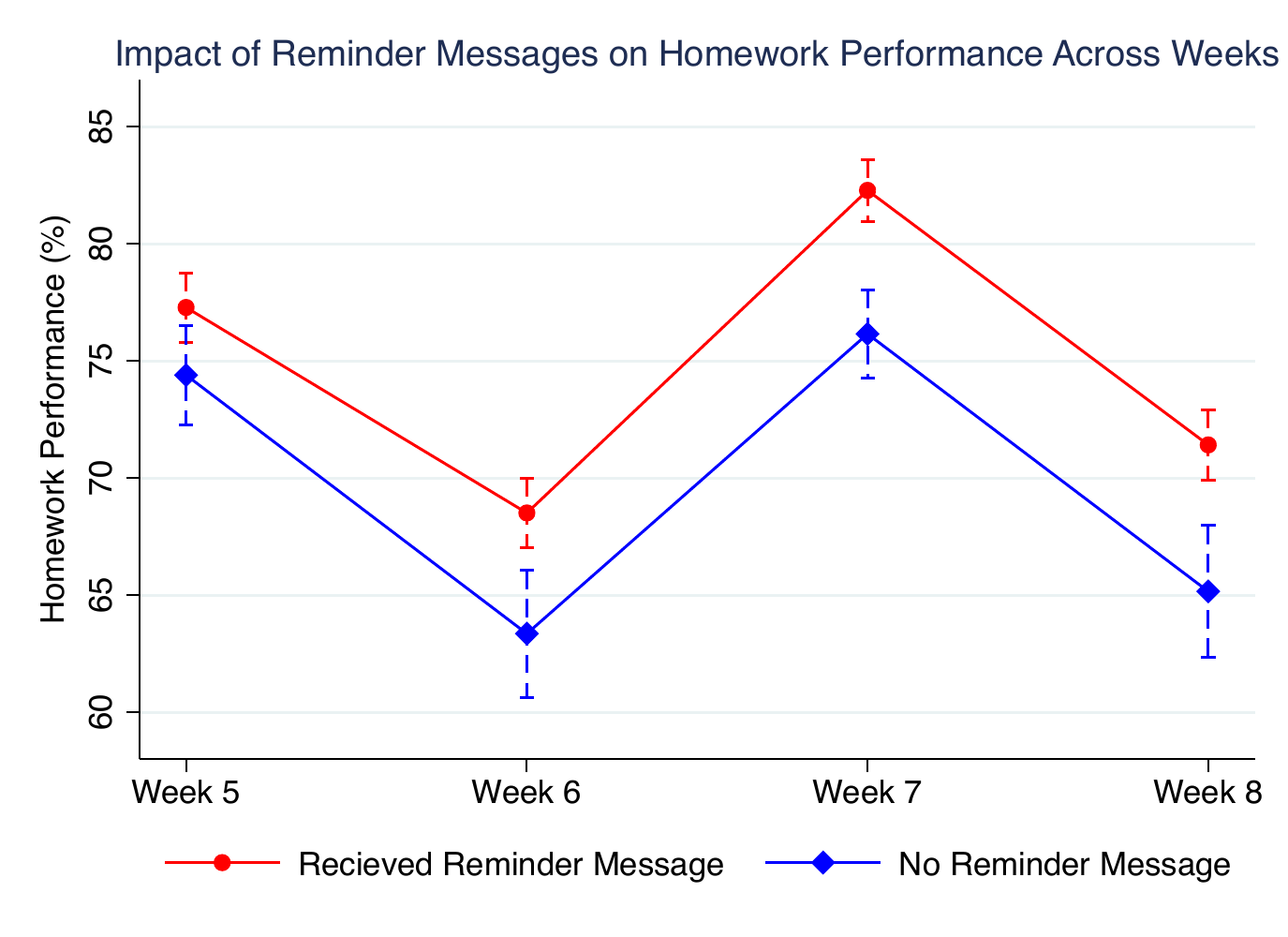}
    \caption{The top figure shows the proportion of students who started the online homework each week from weeks 5 - 8 for students who received a reminder message (circle) and those who did not (diamond). The bottom figure shows the average homework performance each week from weeks 5 - 8 for students who received a reminder message (circle) and those who did not (diamond). The dotted interval at each point indicates the standard error of the mean.}
    \label{fig:attempt_across_weeks}
\end{figure}

\ignore{
\begin{figure}[htb]
    \centering
    \includegraphics[scale=0.5]{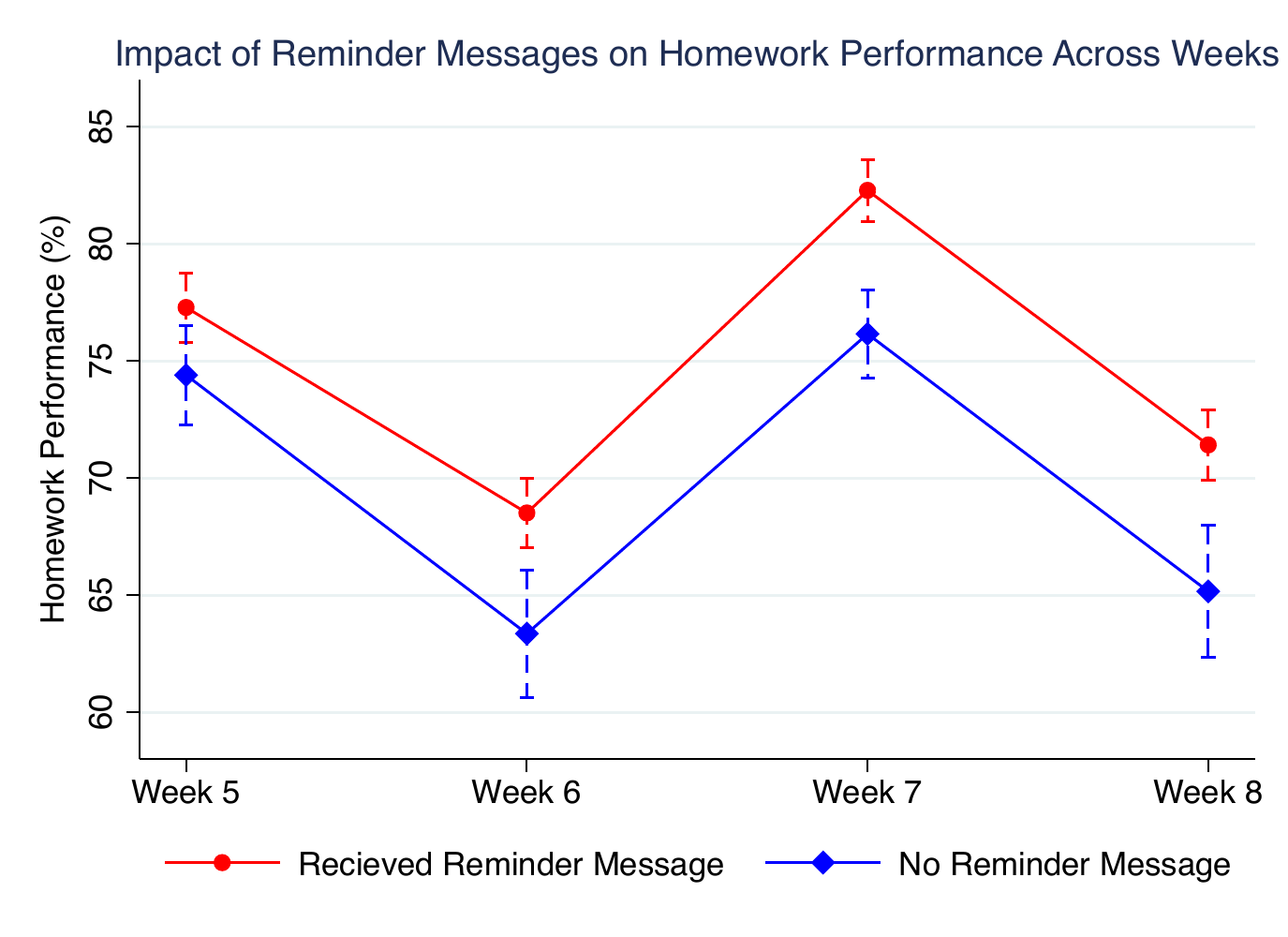}
    \caption{Connected line plot that shows the average homework performance each week from weeks 5 - 8 for students who received a reminder message (circle) and those who did not (diamond). The dotted interval at each point indicates the standard error of the mean.}
    \label{fig:performance_across_weeks}
\end{figure}
}

The messaging intervention had an impact in starting homework. Students who received the reminder messages consistently attempted homework at a higher rate than students who did not as shown by Figure \ref{fig:attempt_across_weeks}. The attempt rate indicates how many students tried the homework, regardless of the number of questions completed or the success. To aggregate the effects of reminder messages across 4 weeks of experimentation we employ a panel data regression model. Table \ref{tab:impact_rate} shows that the reminder messages increased the number of students who started the homework by around 4 percentage points (coefficient = 0.0319; \textit{z} = 2.86; \textit{p} < .01; panel regression with time and participant effects). 

The increase of attempt rate is reflected in the homework performance rates shown in Figure \ref{fig:attempt_across_weeks}. Students who received the reminder messages consistently performed better than students in the control group by attempting or completing more homework problems. We employ a panel data regression model to aggregate the weekly effects. In Table \ref{tab:impact_rate}, we observe that the reminder messages increased the homework average by 4 percentage points (coefficient = 0.0388; \textit{z} = 2.80; \textit{p} < .005; panel regression with time effects and participant effects).

Although the reminder messages impacted students' propensity to start the homework, their impact on students' start time was more limited. As shown in Table \ref{tab:impact_rate}, receiving email reminders do not result in a statistically significant impact on start time (coefficient = 1.105; \textit{z} = 0.86; \textit{p} > .10; panel regression with time and participant effects) or on end time (coefficient = -0.118; \textit{z} = -0.11; \textit{p} > .10; panel regression with time and participant effects).

\subsubsection{Impact of the Reminder Design on Homework Performance}


\begin{figure*}[htb]
    \centering
    \includegraphics[scale=0.39]{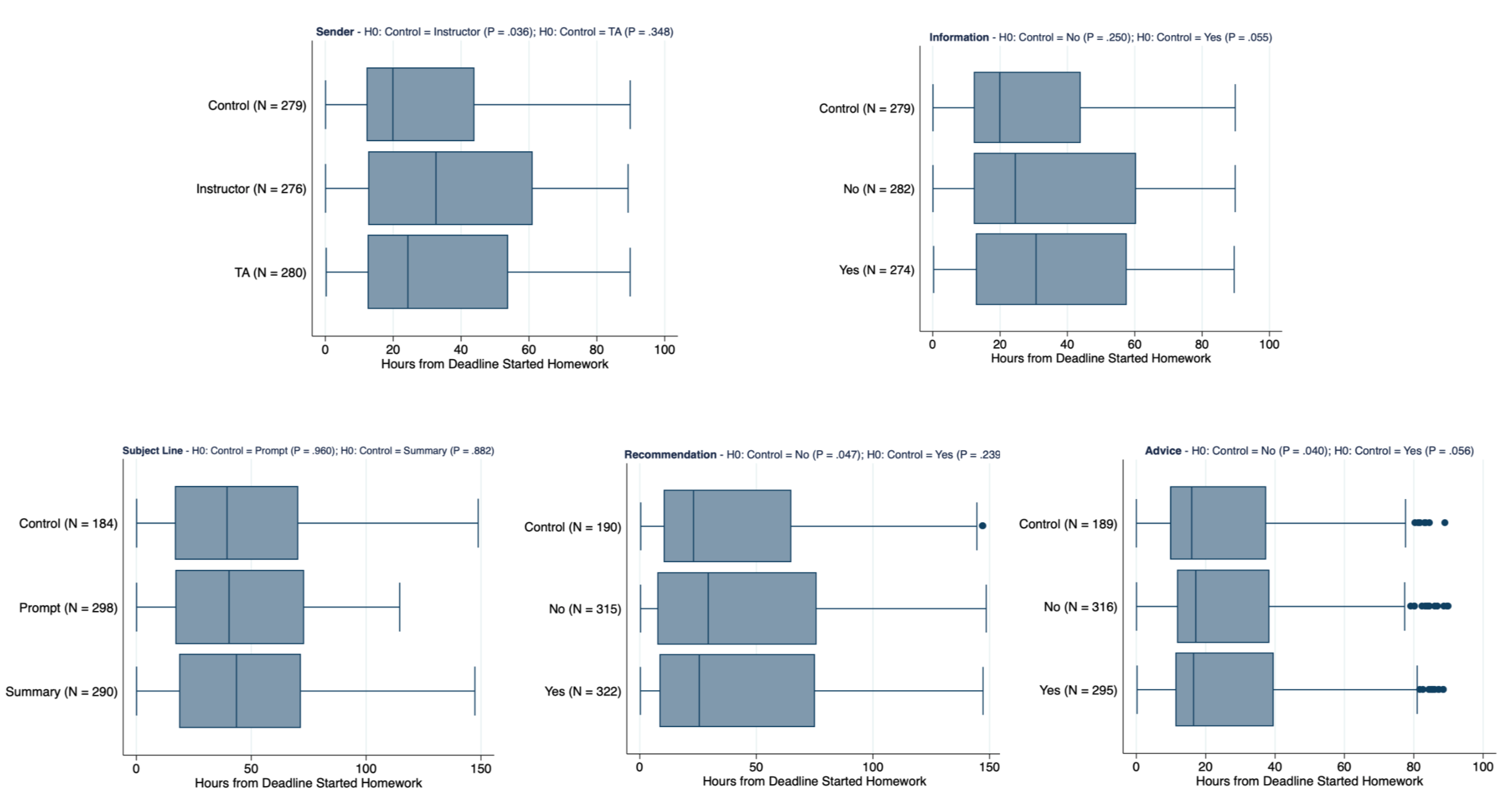}
    \caption{Box plots of student homework start time for conditions in each variation. \textit{p}-values are derived from the Komogorov-Smirnov equality of distributions test, in each variation, compares the distribution of homework start time to the control group. Students who did not attempt the homework were removed from this analysis.}
    \label{fig:box-start}
\end{figure*}

We want to address the effect of each design variation on student action rate. As noted in Figure \ref{fig:box-start}, the distribution of homework start times shows that students tend to start closer to the deadline, thereby creating a skewed start time distribution. As a result, the mean start time and statistical tests based on the mean are not wholly representative of the effect that the emails conditions have. We chose to use box plots to visualize the median and quartiles of start times between conditions and the control group. Furthermore, start time distributions of control-condition within each variation were compared using the Komogorov-Smirnov test. Significant \textit{p}-values for a control-condition comparison means students' start times between the conditions and control are different; in other words, the reminder has an effect, even if not immediately apparent. 

Our results show a significant (\textit{p}<.05) difference in start time distribution for sender (control vs. instructor), recommendation (control vs. no prompt), and advice (control vs. no prompt) variations. Marginal significance (.05$\leq{p}$<.1) was observed for information (control vs. showing the prompt), and advice (control vs. showing the prompt) variations. We did not observe statistical significance for start time distributions for any of the subject line conditions. 

The box plots show an approximate 10-hour increase in median start time in the sender (control vs. instructor), information (control vs. yes - showing the prompt), and recommendation (control vs. no prompt) variations. Smaller increases in median start time are seen the sender (control vs. teaching assistant), information (control vs. no prompt), subject line (control vs. summary), recommendation (control vs. yes - showing the prompt), and both advice conditions over control. We note the median start time for the control group was lower or at least as low as conditions in all cases. 

\begin{table*}[]
\label{tab:my-table}
\begin{tabular}{@{}lllllllllll@{}}
\toprule
Comparison    & Estimate          & $P$-Value          & Estimate             & $P$-Value            & Estimate   & \multicolumn{1}{l|}{$P$-Value}  & Estimate     & \multicolumn{1}{l|}{$P$-Value} & Estimate            & $P$-Value        \\ \midrule
               & \multicolumn{2}{l}{Sender-instructor} & \multicolumn{2}{l}{Information} & \multicolumn{2}{l}{Subject line-prompt} & \multicolumn{2}{l}{Recommendation}            & \multicolumn{2}{l}{Advice} \\
No - Control & 0.003             & 0.988              & 0.0207               & 0.419                & 0.041      & .211                            & $0.081^{**}$ & 0.003                          & 0.029               & .338             \\
Yes - Control & 0.015             & 0.538              & -0.004               & 0.853                & 0.047      & .148                            & $0.093^{**}$ & 0.001                          & $0.081^{**}$        & .008             \\ \bottomrule
\end{tabular}
\caption{The table presents differences in propensity to start the homework for each reminder message variation relative to the control group who did not receive the message. The corresponding p-values are obtained from two independent samples proportions z-test.}
\label{table: attempt_rate}
\end{table*}

Due to the definition of a student attempting a homework problem as a Bernoulli variable, the mean and subsequent \textit{z}-tests in Table \ref{table: attempt_rate} represent the impact of emails on student attempt rate. We note statistically significant (\textit{p}<.05) attempt rate increases for all conditions of recommendation over control (est. 8.1\% no prompt/9.3\% showing the prompt) and  advice (control vs. showing the prompt, est. 8.1\%).

\subsection{Qualitative Results}

\subsubsection{Students' Perceptions of Reminder Messages}

Of the 935 students who responded to the Likert-scale survey questions, most students found it motivating to read the evidence about the benefits of starting the homework early (63\%) and instructor recommendations on how to start early (63\%). Moreover, the majority of students seemed to appreciate that the message included a link to the homework (56\%). We captured the reasons behind the students’ attitudes towards the reminder messages from an open-ended survey question asking about their initial reaction and if it affected their behaviours. When asked the question, \textit{``After receiving the message, did it motivate you to set aside time to work on [homework], or did it change your behaviour at all?''}, 474 students provided an answer, with most of them indicating a positive behaviour change after receiving the reminder (57\%). About half of students who indicated that the messages were motivating (27\% of the total) did not provide a reason for finding the messages motivational. Of those who responded, the majority (20\% of the total) shared that the messages helped them reduce procrastination by starting early and making a plan: \textit{``The message made me consider my time ahead, and plan when to do the [homework] questions.''} 

Of the students who indicated that the reminder messages did not influence a behaviour change, most (28\% of the total) did not explain why. The majority of those who provided a reason noted that they already had a plan (7.5\%) or had already completed the work (2.9\%). For those students, the messages could be frustrating: \textit{``I was a little annoyed because I have already started doing [homework] for that week''} and \textit{``it becomes annoying ... I know I will do it before the deadline.''} 
For other students, the message was appreciated and could potentially lead to helping them in some contexts (13.6\% of the total), but they noted that other factors got in the way of inducing a change: \textit{``Yes, but when I saw it I was busy and later on forgot about it.''}

\textbf{Students' Emotional or Attitudinal Responses to Emails}. 

Whether or not the emails were effective, they often produced an emotional response, which suggests that emails need to be targeted carefully. Students who had not yet started sometimes felt that they were being cared for: \textit{``I felt like the [course code] team actually cares about me and my learning that it surprised me.''} (1.4\% of the total), but if they were unable to start or were struggling in the course, the extra attention could be stressful: \textit{``if anything it kind of stressed me out knowing I have something due and I haven't done it yet''} (6\% of the total). Negative reactions were more common from students who had already completed the work (4\% of the total). The messages caused them to question whether their work had been submitted correctly: \textit{``It made me panic not thinking I'd turned in [the] assignments.''} 

We observed similar patterns in the interviews. Some students (6/9) reported in the interviews that they liked receiving the reminder messages as they were helpful for spurring homework completion. One student shared: \textit{``the instructor would send us emails ... and those were so helpful for me because I procrastinate a lot, so getting reminders by one of my favorite professors was actually so much fun for me.''}. The remaining 3/9 students had neutral opinions about the reminders. A few students (2/9) mentioned that having a notification on their phones was a positive aspect of the reminders, especially if the due date was in the header of the message.

Students (5/9) mentioned in the interviews that they appreciated the messages that included a recommendation about how to start early as they agreed that it motivated them to make a study plan or to consider starting early, which is the goal of the implementation intentions principle. One student highlighted that they were able to apply the recommended actions to other courses: \textit{``That piece of advice in that email actually helped me to like go to other classes, and like, read my other assignments before starting them. So that was really helpful for me and I learned from that because now I've actually like apply that for my other classes''}.

Students (2/9) said that the reminder messages gave them the feeling of being valued and cared for by the course team: \textit{``I planned everything beforehand. So it was really nice to receive this because most courses don't even tell you at all. There's no prompt, there's nothing. ... I think there's a lot more intention and care.''}. One student shared their reaction to the first email they received from the course instructor: \textit{``It seems like [the instructor] genuinely wanted to like do his best to help you learn and be there for you make sure that you know that there are resources that you can use and that you can reach out to him. He was a very caring professor and it seemed that this was his way of helping us. When I received the first email, it made me happy because you don't usually get those from professors''}.

\textbf{Student perceptions of the negative effects of reminder messages}.

Some comments from students' surveys and interviews about the reminder emails highlighted design aspects that should be considered for future work. For instance, students mentioned that the reminder message looked like a spam email. 
Including the students' names brings up a sense of personalization that is appreciated by some students, but it might cause confusion and the feeling of getting targeted: \textit{``When I saw my name ... I panicked thinking I did something wrong. Then I realized it was sent to everyone.''} Several students commented that they replied to the reminder message providing their plan to start early, but they did not get a response which caused discomfort: \textit{``It stressed me out beyond belief! Also, nobody replied to them?''}. Students were prompted to reply to the email to motivate them to answer. 

During the interviews, some students (3/9) mentioned that they consistently received the reminder messages after completing their homework, and felt patronized by the reminder. One interviewee said: \textit{``I always got [the email] after I finished everything...it was kind of like annoying. Cause I’m already done, why are you still nagging me about this?''} However, all students in the interviews who reported receiving the reminders after completing their homework mentioned that they categorize the reminders as beneficial for students who struggle with time management when asked about their overall opinions of the emails.

\begin{table*}[htbp]\centering

\begin{tabular}{l|c c c c c}
            &\multicolumn{1}{c}{(1)}&\multicolumn{1}{c}{(2)}&\multicolumn{1}{c}{(3)}&\multicolumn{1}{c}{(4)}&\multicolumn{1}{c}{(5)}\\
            & Email & Better condition & Email Attention & Student Action & Instructor Confidence\\
            & Variation & Prediction & Rate Difference & Rate Difference & (-5 to +5)\\
\hline
Instructor 1& Control vs. Email &    Email        & -         &         10\%        &         +2  \\
Instructor 2& (Week 5-8) &    Email         & -         &         10\%        &         +4 \\
\hline
Instructor 1& Sender &    Instructor          & 5\%         &         10\%        &         +2  \\
Instructor 2& (Week 5) &    Instructor        & unsure         &      5\%        &         +1 \\
\hline
Instructor 1& Information &    Include information       & 0\%         &         5\%        &         0  \\
Instructor 2& (Week 5) &    Include information         & 10\%         &         5\%        &         +2 \\
\hline
Instructor 1& Subject Line &    Deadline summary        & 0\%         &         20\%        &         +2 \\
Instructor 2& (Week 6) &    Deadline summary          & 10\%         &         5\%        &         +2 \\
\hline
Instructor 1& Recommendation &    No recommendation       & 0\%         &         5\%        &         -2  \\
Instructor 2& (Week 7) &    No recommendation & 10\%         &         5\%        &         0 \\
\hline
Instructor 1& Advice &    Include advice         & 0\%         &         5\%        &         -1  \\
Instructor 2& (Week 8) &    Include advice       & 10\%         &         5\%        &         +2 \\
\end{tabular}
\caption{Summary of instructor survey responses. Email Variation represents the email design factor the instructors  considered. Better condition Prediction is the condition that each instructor predicted was best for getting students to start earlier and take action. Email Attention Rate Difference represents each instructor's prediction of student engagement increase given better condition. Student Action Rate Difference is each instructor's prediction of student action increase (doing their weekly homework) for the better condition. Instructor Confidence is how confident they were in their predictions, where -5 is not confident at all and +5 is extremely confident. Refer to Section 3.3 for the design of each condition.}
\vspace*{-2em}
\label{tab:instructor}
\end{table*}

\subsubsection{Instructor Predictions on Reminder Messages}
We summarized the data collected from the two instructor surveys in Table \ref{tab:instructor}. These surveys help visualize the differences between the effect observed and the predictions. For all variations, both instructors made the same prediction for which condition would be better. Instructors predicted a $\leq$10\% increase for all predictions, with Instructor 2 being unsure of how sender would affect engagement. Instructors also generally predicted a $\leq$10\% increase in students taking action to do weekly homework given the better condition, the only exception being Instructor 1, who emphasized that including the deadline in a subject line would have a large (20\%) impact. Instructor 1 was moderately confident when predicting variations outside the email body: Control vs. Treatment (+2), Sender(+2), Subject Line (+2), while less confident when predicting conditions in the email body: Information (0), Recommendation (-2), Advice (-1). Instructor 2 was moderately confident in their predictions (+2 or +4) for all variations, except for Sender (+1/expressed uncertainty) and Recommendation (0). 

\subsubsection{Instructor Intuition Relative to Quantitative Results}
Based on the quantitative results, instructors can generally correctly predict the better condition for students to start early (corresponding to start time Figure \ref{fig:box-start}) and take action (corresponding to attempt rate Table \ref{table: attempt_rate}) and aggregated overall in Table \ref{tab:impact_rate}. However, instructors  overestimated the effectiveness of some conditions. For example, the overall effectiveness of emails in Table \ref{tab:impact_rate} was <5\% for attempt rate (statistically significant) and <2 hour increase for start time (non-statistically significant), but both instructors predicted a 10\% action ate difference. We again note instructors are also not wholly confident in all their predictions, even if they did consistently pick the better condition.

\section{Discussion}

We explored alternative designs for using emails to prompt students to plan and stay on track with their homework. We show how we can use randomized A/B comparisons to test which ideas might impact behaviour. We present the particular dependent variables and metrics anyone can use to conduct such studies - start time and attempt rate. We also explore  the design space of factors to test, which we hope can inspire a wide range of follow-up studies using a similar paradigm. Our results illustrate some of the tradeoffs instructors might consider, as well as empirical insights into what may be more or less effective, to guide the necessary future studies needed for expanding and generalizing this work to other contexts. The randomized A/B comparisons investigated whether and how different reflection/reminder messages might motivate students to complete their homework and start earlier. Procrastination and homework performance are key issues that instructors in online learning contexts need to address, and we provide a feasible intervention.

Some instructors were skeptical that students would change their behavior from a simple message – or even pay attention to them. However, identifying reminder messages that \textit{can} shift behavior is extremely valuable because it is a low-cost and widely applicable intervention, and one of the major ways instructors are reaching out to students outside of class time. Interestingly, we found that the emails prompting students to reflect on how they could start earlier did not increase how many hours before the due date students began.  However, it did increase the number of students who actually started the homework (as many students in a week never started the homework) and, as a result, increased the average performance on the homework.

We used psychologically-grounded principles to create the emails in a way that they are helpful and encouraging to students. One of the psychological theories that we used to design these reminder messages and make them more motivating is the implementation intentions strategy \cite{gollwitzer1999implementation, klassen2008academic}, where students were asked to reflect on a plan with information on when and how to improve to transform good intentions into behavioural change. Furthermore, we also used practices motivated by self-efficacy and sharing mastery experiences with students. 

We tried to understand students' experiences and perspectives on receiving the emails. We collected a total of 935 survey responses, and combined this with interviews with 9 students, to get converging perspectives of large-scale data and in-depth discussions. The overall positive impact of the reminder emails on students’ attempts of homework aligns with other research on email interventions \cite{lim2021impact, nikolayeva2020does, yeomans2017planning}. However, it is interesting that the intervention did not significantly lead students to start early. A possible explanation for this can be gleaned from the qualitative response of students - at least one student expressed that they did not receive feedback on their plan to start the homework early, with a suggestion that it affected their motivation. Possibly, a more dialogic or coaching approach might further encourage students to work out their plan, but this needs to be investigated. 

For students who were less impacted by the intervention, i.e., did not attempt the homework even after reminders, it is possible that they believed that they were already self-regulating effectively, or did not feel confident about doing the homework. It may also be the case that these students are facing some issues that might need to be addressed in a different way, other than through emails. More information is needed about students who are not engaging or responding to the interventions. Future research can investigate students’ baseline self-regulated learning habits or self-efficacy, to obtain further insights into how email interventions could be effectively tailored to them.

The qualitative data collected from students sheds light on when and why such reminder messages can be seen as valuable by students versus annoying or excessive. The majority of students said they found the message helpful, and some felt that the intervention showed that the instructional team cared about their learning. Despite this, our quantitative metrics suggested only a fraction of those actually changed their behaviour, suggesting how future work can go deeper into understanding how exactly to craft the prompts to tackle the mechanisms of behaviour change. The qualitative finding around students’ appreciation of instructor care as demonstrated through the emails resonates with other research such as \cite{lim2021impact, arthars2019empowering}, showing the affordances of technology-mediated communication can foster a feeling of greater support in students. On the other hand, some students were stressed by a reminder about doing homework, and a few expressed irritation because they already had a plan for completing the homework or were already done. 

In terms of how useful the results were in testing predictions from instructors, we found that though instructors were generally able to predict the best condition to send in an email (though not always confidently), they would often overestimate the effectiveness of the emails when translated into student attempt rate and start times. This reinforces the notion that involving instructor intuition in the design process of reminder emails would allow for better conditions to be picked \cite{motz2018causal}. At the same time, this demonstrates the need for data from real-life A/B comparison experiments to help instructors quantify their intuitions and better understand the effectiveness of their communications.

We found suggestive evidence some design choices motivated students to start earlier compared to other conditions during certain weeks such as emails from instructors, including information about starting early, and not including recommendations or advice prompts. Similarly, we also found evidence which suggests some design choices can help significantly increase student attempt rate such as including an advice prompt. In addition, students receiving certain conditions would significantly alter their start time distributions, indicating our design choices had an impact even if not directly summarized by the mean or median. Finally, on an aggregate level over the period of 4 weeks, it was more beneficial for students to receive a reminder email than not in terms of doing their homework and starting it early.

These results can guide instructors in understanding students' context and decision making, In the time of COVID-19, with channels like emails, instructors should learn how to better reach students, and about how to balance the benefit to some students with the inconvenience to others. 


\section{Limitations \& Future Work}

Students who found the reminder messages not helpful or motivating still received the messages, which can cause a negative effect as they can feel annoyed and stressed. To help students who find the reminder messages helpful, but at the same time, consider the students who do not find value in receiving these, we would like to use more adaptive and personalized messages in future deployments. If there were a closer integration between the homework system and the learning management system, we could provide more customization and not send out a message if the student has already started the homework. Moreover, obtaining more information about students’ prior knowledge and motivations may provide possible targets for personalization, which may influence their response to the interventions. For the students who plan their week ahead but would like to receive reminders, we could provide an option for changing the reminder messages’ time and frequency. Students who found that the messages did not fit their schedule could schedule one or multiple reminders based on when they wish to receive them, as these messages contain more than just reminders but also prompts that can motivate them to take action.  

Students were aware of the format of the reminder messages after a certain number of occurrences, which could have led them to stop opening and reading these emails. If we chose to add more personalized information to the emails, such as current marks and performance rates \cite{edwards2015examining}, we could make the messages more novel and varied, to reduce the perception that the messages are formulaic and to attract readers. The question becomes if the reminder function is sufficient for students, and having a message attached to it might encourage students to pay attention to it.

We could not directly measure student engagement with the emails at the time of sending such as if and when they opened the emails, though we could get parts of this data through later interviews. Student engagement is important in making sure they get the full benefit of the reminder emails, while also ensuring the effect between design decision and student action are accurately inferred.

Future evaluation that could build on our work includes: (1) Exploring how these findings generalize to contexts beyond CS1; (2) Exploring in more depth the characteristics of students who appreciated the reminders and those who had negative opinions as the qualitative responses were mixed; (3) Investigating improvements on these reminder messages, such as alternative ways of getting students to start earlier, and emphasizing the importance of students planning how to complete their homework after starting; (4) Introducing additional measures of student engagement with the emails; (5) Furthering evidence through A/B comparisons on effective conditions independent of the week they were sent. Although sending emails to students is a ubiquitous everyday activity, it is important for instructors and researchers to understand how to use widely used communication channels to motivate and reach students, especially during a pandemic, and which of these observable measures of behaviour are effective.

\section{Conclusion}

This paper leverages time-stamped homework submissions, qualitative data from nearly 1000 students and two course instructors, and randomized A/B comparisons to assess the effectiveness of reminder messages in getting students to start their online homework early. We find evidence, over four weeks, to suggest that reminder messages have a modest effect in promoting more students to start their homework, but are not as observably effective in altering the time at which they start the homework. We also found suggestive evidence of effective email design conditions which could be further explored. Combining instructor predictions about email design variations, we found although instructor intuition is important, they might not always be able to confidently quantify this intuition, and using experimental A/B comparisons can help them test their ideas. Furthermore, from surveying and interviewing students we discover some negative aspects of the reminders such that students may find the email motivating but are unsure on how to make progress, some students can become further stressed being told about upcoming deadlines, and some students may become annoyed if they have already made significant progress on the homework. These trends in student perspective align with our experimental findings that reminder messages are helpful to a subset of students that are encouraged to attempt and start the homework due to the reminder, but they also provide insight into how to improve our reminders.



\section{Acknowledgements}
This work was partially supported by the Natural Sciences and Engineering Research Council of Canada (NSERC) (No. RGPIN-2019-06968), as well as by the Office of Naval Research (ONR) (No. N00014-21-1-2576)

\bibliographystyle{ACM-Reference-Format}
\bibliography{references}

\end{document}